\documentclass[prl,twocolumn,aps,amsmath,nofootinbib,superscriptaddress]{revtex4}

\usepackage{graphicx}
\usepackage{bm}
\usepackage{epsfig}

  \newcount\hour \newcount\minute
  \hour=\time \divide \hour by 60
  \minute=\time
  \newcommand{\mydate}{\ \today \ - \number\hour :\ifnum \minute<10 0\fi 
\number\minute}
\def\nslash{n\!\!\!\slash}
\def\bnslash{\bar n\!\!\!\slash}

\def\OMIT#1{}

\newcommand{\nn}{\nonumber} 

\newcommand{\bn}{{\bar n}}
\newcommand{\bea}{\begin{eqnarray}}
\newcommand{\eea}{\end{eqnarray}}

\newcommand{\bnP}{\bar {\cal P}}

\newcommand{\cP}{{\cal P}}
\newcommand{\cPslash}{ {\cal P}\!\!\!\!\slash}

\newcommand{\mcdot}{\!\cdot\!}

\newcommand{\SCETa}{\mbox{${\rm SCET}_{\rm I}$ }}
\newcommand{\SCETb}{\mbox{${\rm SCET}_{\rm II}$ }}

\begin{document}

%%%%%%%%%%%%%%%%%%%%%%%%%%%%%%%%%%%%%%%%%%
%Define Title, Author, Address, Preprint#

\preprint{ \hbox{MIT-CTP 3469} \hbox{CMU-HEP-04-01} \hbox{CALT-68-2475}
 \hbox{hep-ph/0401188}  }

\title{\boldmath 
$B \to M_1 M_2$: Factorization, Charming Penguins, Strong Phases, and
  Polarization \\
%\ \mydate 
}

\author{Christian W.~Bauer}
\affiliation{California Institute of Technology, Pasadena, CA 91125}
\author{Dan Pirjol}
\affiliation{Center for Theoretical Physics, Massachusetts Institute of
  Technology, Cambridge, MA 02139}
\author{Ira Z.~Rothstein}
\affiliation{Department of Physics, Carnegie Mellon University,
    Pittsburgh, PA 15213  \vspace{0.3cm}}
\author{Iain W. Stewart\vspace{0.4cm}}
\affiliation{Center for Theoretical Physics, Massachusetts Institute of
  Technology, Cambridge, MA 02139}
%%%%%%%%%%%%%%%%%%%%%%%%%%%%%%%%%%%%%%%%%%
\begin{abstract}
  
  Using the soft-collinear effective theory we derive the factorization theorem
  for the decays $B\to M_1 M_2$ with $M_{1,2}=\pi,K,\rho,K^*$, at leading order
  in $\Lambda/E_{M}$ and $\Lambda/m_b$. The results derived here apply even if
  $\alpha_s(E_M \Lambda)$ is not perturbative, and we prove that the physics
  sensitive to the $E\Lambda$ scale is the same in $B\to M_1 M_2$ and $B\to M$
  form factors.  We argue that $c\bar c$ penguins could give long-distance
  effects at leading order. Decays to two transversely polarized vector mesons
  are discussed.  Analyzing $B\to\pi\pi$ we find predictions for
  $B^0\to\pi^0\pi^0$ and $|V_{ub}| f_+^{B\to \pi}(0)$ as a function of $\gamma$.

\end{abstract}

\maketitle
 
Decays of $B$ mesons to two light mesons are important for the study of $CP$
violation in the standard model.  In~\cite{BBNS} it was suggested that since
$m_b, E_M \gg\Lambda, m_{M}$ the amplitudes should factorize into simpler
non-perturbative objects, and the proposed factorization theorem was checked at
one-loop. This approach is often referred to as ``QCD Factorization'' (QCDF).
Factorization has also been considered in the ``perturbative QCD'' (pQCD)
approach~\cite{Keum}.  These approaches rely on a perturbative expansion in
$\alpha_s(E_M \Lambda)$.  The results obtained from factorization are quite
predictive and may allow us to answer fundamental questions about the standard
model. At the current time several important issues remain to be answered. These
include: 1) The extent to which the results are model independent consequences
of QCD (since QCD is a predictive theory any model independent limit must give
the same answer in different approaches). A complete proof of a factorization
theorem will answer this question.  2) Unambiguous definitions of any
nonperturbative hadronic parameters which appear are required. This allows the
universality of parameters to be understood, as well as making clear the extent
to which predictions rely on model dependent assumptions about parameter values.
3) Does the power expansion converge?  If power suppressed contributions really
compete with leading order contributions as some studies~\cite{BBNS2,Ciuchini2}
suggest then the expansion can not be trusted. In this case the only hope is a
systematic modification of the power counting to promote these effects to
leading order, or an identification of certain observables that are free from
this problem.

The soft collinear effective theory (SCET) \cite{SCET,bfprs} provides the
necessary tools to address these issues. A first study of SCET factorization for
$B\to \pi\pi$ has been made in~\cite{chay}.  In this paper we go beyond
Refs.~\cite{BBNS,Keum,chay} in several ways. We first reduce the SCET operator
basis to its minimal form and extend it to allow for all $B\to M_1 M_2$ decays
(including two vectors).  Our results show that all of the so-called ``hard
spectator'' contributions are already present in the form factors, just with
different hard Wilson coefficients.  We also derive a form of the factorization
theorem which does not rely on a perturbative expansion in $\alpha_s(E_M
\Lambda)$, and show that the non-perturbative parameters are still the same as
those in the $B\to M$ form factors.  In our analysis long distance $c\bar c$
penguins~\cite{Ciuchini,Colangelo} are investigated, but are left unfactorized.
For natures values of $m_b$ and $m_c$ we give an argument why these
contributions can be leading order. This is contrary to expectations that they
are power suppressed~\cite{BBNS}, but in agreement with expectations
in~\cite{Ciuchini,Colangelo,Ciuchini2}.  The presence of these contributions
could introduce large LO non-perturbative strong phases. Even in observables
that are free from charming penguins our results differ phenomenologically from
Ref.~\cite{BBNS}.  In particular while the power counting in Ref.~\cite{BBNS}
requires a heirarchy in parameters $\zeta_J^{B\pi}\ll \zeta^{B\pi}$, we show
that SCET allows for other possibilities such as $\zeta_J^{B\pi}\sim
\zeta^{B\pi}$.  We demonstrate that the LO SCET results are in agreement with
current $B\to\pi\pi$ data, and find current central values favor $\zeta_J^{B\pi}
\gtrsim \zeta^{B\pi}$, albeit with fairly large uncertainties.

We set $M=P$ when discussing pseudoscalars, $M=V$ for vectors, and use an $M$ to
denote either. The decays $B \to M_1 M_2$ are mediated in full QCD by the weak
$\Delta B=1$ Hamiltonian, which for $\Delta S=0$ reads
\begin{eqnarray} \label{Hw}
 H_W = \frac{G_F}{\sqrt{2}} \sum_{p=u,c} \lambda_p^{(d)}
 \Big( C_1 O_1^p + C_2 O_2^p 
  +\!\!\! \sum_{i=3}^{10,7\gamma,8g}\!\! C_i O_i \Big),
\end{eqnarray}
where the CKM factor is $\lambda_p^{(f)} = V_{pb} V^*_{pf}$ with $f=d$.  The
standard basis of $f=d$ operators are (with $O^p_{1}\leftrightarrow O^p_{2}$
relative to~\cite{fullWilson})
\begin{eqnarray}\label{fullops}
 O_1^p \!\! &=&\!\! (\overline{p} b)_{V\!-\!A}
  (\overline{d} p)_{V\!-\!A}, \ \
 O_2^p = (\overline{p}_{\beta} b_{\alpha})_{V\!-\!A}
  (\overline{d}_{\alpha} p_{\beta})_{V\!-\!A}, \nonumber \\
 O_{3,4} \!\! &=& \!\! \big\{ (\overline{d} b)_{V\!-\!A}
  (\overline{q} q)_{V\! - \!A}\,, (\overline{d}_{\beta} b_{\alpha})_{V\!-\!A}
  (\overline{q}_{\alpha} q_{\beta})_{V\! - \!A} \big\}, \nonumber \\
 O_{5,6} \!\! &=& \!\! \big\{ (\overline{d} b)_{V\!-\!A}
  (\overline{q} q)_{V\! + \!A}\,, (\overline{d}_{\beta} b_{\alpha})_{V\!-\!A}
  (\overline{q}_{\alpha} q_{\beta})_{V\! + \!A} \big\}, \nonumber \\
 O_{7,8} \!\! &=& \frac{3e_q}{2}\!\! \big\{ (\overline{d} b)_{V\!-\!A}
  (\overline{q} q)_{V\! + \!A}\,, (\overline{d}_{\beta} b_{\alpha})_{V\!-\!A}
  (\overline{q}_{\alpha} q_{\beta})_{V\! + \!A} \big\}, \nonumber \\
 O_{9,10} \!\! &=& \frac{3e_q}{2}\!\! \big\{ (\overline{d} b)_{V\!-\!A}
  (\overline{q} q)_{V\! - \!A}\,, (\overline{d}_{\beta} b_{\alpha})_{V\!-\!A}
  (\overline{q}_{\alpha} q_{\beta})_{V\! - \!A} \big\}, \nonumber \\
 O_{7\gamma,8g} \!\! &=&\!\!  -\frac{m_b}{8\pi^2}\ \overline{d}\, \sigma^{\mu\nu}
  \{e F_{\mu\nu},g G_{\mu\nu}^a T^a\} (1\!+\! \gamma_5)  b \,.
\end{eqnarray}
Here the sum over $q=u,d,s,c,b$ is implicit, $\alpha, \beta$ are color indices
and $e_q$ are electric charges. The $\Delta S=1$ $H_W$ is obtained by replacing
$(f=d)\to (f=s)$ in Eqs.~(\ref{Hw},\ref{fullops}).  The coefficients in
Eq.~(\ref{Hw}) are known at NLL order~\cite{fullWilson}. In the NDR scheme
taking $\alpha_s(m_Z)=0.118$ and $m_b=4.8\,{\rm GeV}$ gives
$C_{7\gamma}(m_b)=-.317$, $C_{8g}(m_b)
=-0.149$ and
\begin{eqnarray}
 && 
C_{1-10}(m_b) = \{
  1.080\,, 
  -.177\,,
  .011\,,
 -.033\,, 
  .010\,, 
 -.040 \,, 
  \nn\\
 && \ \ 
  4.9 \!\times\! 10^{-4} \,,
  4.6 \!\times\! 10^{-4} \,,
  -9.8 \!\times\! 10^{-3} \,,
  1.9 \!\times\! 10^{-3} \} \,.
\end{eqnarray}

\begin{figure}[t!]
  \centerline{ \mbox{\epsfysize=2truecm \hbox{\epsfbox{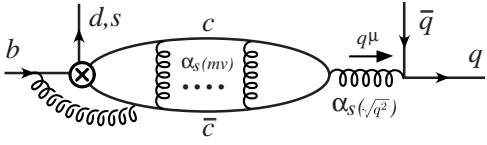}} } }
  \vskip-0.3cm 
  {\caption[1]{Example of long distance charming penguins. The $mv$ gluons
      are nonperturbative and LO soft gluons are exchanged by the $b$, $c$, $\bar
      c$ and spectator quark which is not shown.  }
\label{fig:cpenguin} }
\vskip -0.2cm
\end{figure}

\begin{figure}[t!]
  \centerline{ \mbox{\epsfysize=2truecm \hbox{\epsfbox{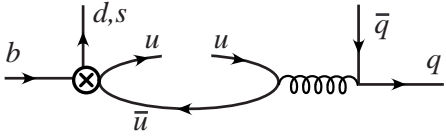}} } }
  \vskip-0.3cm 
  {\caption[1]{Example of a long distance light quark penguin which matches onto
      a power suppressed operator. The $\overline q$ goes in the $\bn$
      direction, the $q$ goes in the $n$ direction, the broken $u$ quark line is
      soft or collinear and the $\bar u$ and gluon remain hard.   }
\label{fig:upenguin} }
\vskip -0.2cm
\end{figure}

The relevant scales in $B\to M_1 M_2$ are $m_b$, $m_c$, the jet scale
$\sqrt{E\Lambda}$ and $\Lambda$. Varying $\Lambda$ between $100-1000\,{\rm MeV}$
the jet scale is numerically in the range $\sqrt{E\Lambda}\simeq 0.5-1.6\,{\rm
  GeV}$. Integrating out $\sim m_b$ fluctuations, the effective Hamiltonian in
\SCETa\cite{bps4} can be written as
\begin{eqnarray} \label{match}
 H_W \!\!&=  &\!\! \frac{2G_F}{\sqrt{2}} \sum _{n,\bn} \bigg\{ 
  \sum_i \int [d\omega_{j}]_{j=1}^{3}
       c_i^{(f)}(\omega_j)  Q_{if}^{(0)}(\omega_j) \nn\\ 
 && \hspace{-1cm}
  + \sum_i \int [d\omega_{j}]_{j=1}^{4}  b^{(f)}_i(\omega_j) 
  Q_{if}^{(1)}(\omega_j) 
  + {\cal Q}_{c\bar c} + \ldots \bigg\} \,,
\end{eqnarray}
where $c_i^{(f)}$ and $b_i^{(f)}$ are Wilson coefficients, the ellipses are
higher order terms in $\Lambda/Q$, $Q=\{m_b,E\}$, and ${\cal Q}_{c\bar c}$
denotes operators appearing in long distance charm effects as in
Fig.~\ref{fig:cpenguin}. Penguin contractions with light quark loops are
included in matching onto ${Q}_{if}^{(0,1)}$ since their long distance
contributions are power suppressed~\cite{BBNS}. The long-distance contributions
occur when one or both of the quark lines in the penguin loop become soft or
collinear. In matching onto SCET these quark lines are left uncontracted and
give rise to higher dimension operators which are power suppressed. An example
which gives rise to a six quark operator is given in Fig.~\ref{fig:upenguin}.

In penguin contractions with charm quarks the situation is different due to the
threshold region. For the $c \bar c$ system the offshellness depends on the
value of $q^2=m_b^2 x$, and long distance contributions from $x\to 0$ or $x\to
1$ are suppressed~\cite{BBNS2}.  However, for $q^2 \sim 4 m_c^2$ the charm quarks
are moving non-relativistically. This region corresponds to momentum fractions
$x \simeq 4 m_c^2/m_b^2 \simeq 0.4$ in the middle of the distribution
$\phi_M(x)$. These contributions have one $\alpha_s(2m_c)$, but can not be
calculated perturbatively.  Using NRQCD power counting they are ``suppressed''
by ${\cal O}(v)$ with $v \simeq 0.4-0.5$.  Thus we conclude that these
contributions may be leading order, and comparable in size to other penguin
terms such as those from the small Wilson coefficients $C_{3-6}$. A rigorous
account of these long distance $c\bar c$ penguin contractions can only be
obtained by deriving a factorization theorem for them, however we do not attempt
to do so here, and therefore do not write down operators for $Q_{c\bar c}$.

\begin{figure}[t!]
 \centerline{
 \mbox{\epsfysize=5truecm \hbox{\epsfbox{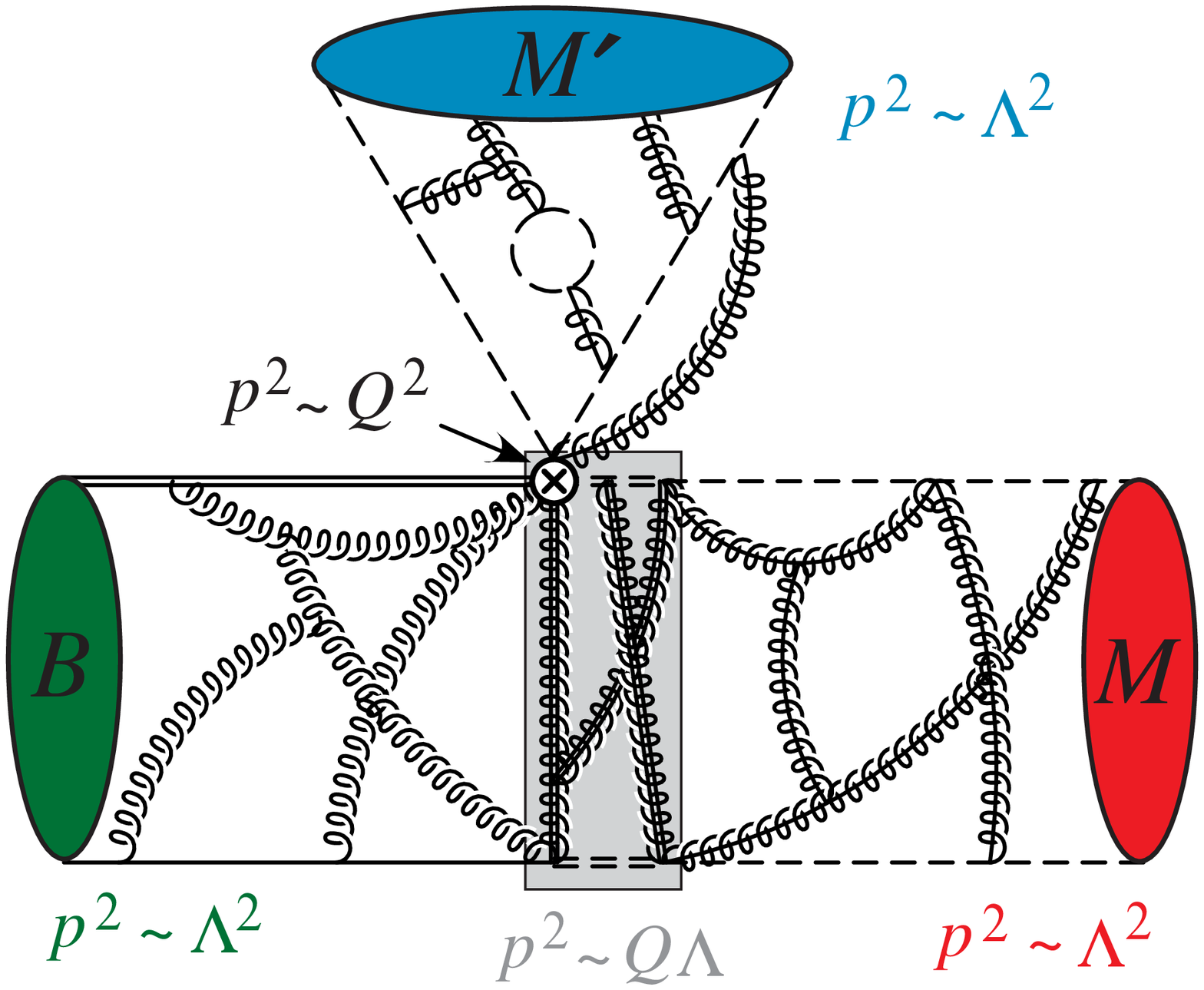}} }
 } 
\vskip -0.2cm
{\caption[1]{Factorization of $B\to M M^{\prime}$ in SCET.}
\label{fig:factor} }
\vskip -0.3cm
\end{figure}
In Eq.~(\ref{match}) the ${\cal O}(\lambda^0)$ operators are [sum over
$q=u,d,s$]
\begin{eqnarray} \label{Q0}
  Q_{1d}^{(0)} &=&  \big[ \bar u_{n,\omega_1} \bnslash P_L b_v\big]
  \big[ \bar d_{\bn,\omega_2}  \nslash P_L u_{\bn,\omega_3} \big]
  \,,  \\
  Q_{2d,3d}^{(0)} &=&  \big[ \bar d_{n,\omega_1} \bnslash P_L b_v \big]
  \big[ \bar u_{\bn,\omega_2} \nslash P_{L,R} u_{\bn,\omega_3} \big]
   \,,\nn \\
  Q_{4d}^{(0)} &=&  \big[ \bar q_{n,\omega_1} \bnslash P_L b_v \big]
  \big[ \bar d_{\bn,\omega_2} \nslash P_{L}\, q_{\bn,\omega_3} \big]
   \,, \nn \\
  Q_{5d,6d}^{(0)} &=&  % \sum_q 
  \big[ \bar d_{n,\omega_1} \bnslash P_L b_v \big]
  \big[ \bar q_{\bn,\omega_2} \nslash P_{L,R} q_{\bn,\omega_3} \big]
  \,, \nn
\end{eqnarray}
with $Q_{is}^{(0)}$ obtained by swapping $\bar d\to \bar s$. In Eq.~(\ref{Q0}) the
``quark'' fields with subscripts $n$ and $\bn$ are products of collinear quark
fields and Wilson lines with large momenta $\omega_i$. For example
\begin{eqnarray}
  \bar u_{n,\omega} = [ \bar\xi_n^{(u)} W_n\, \delta(\omega\!-\!
\bn\mcdot\cP^\dagger) ]\,,
\end{eqnarray}
where $\bar\xi_n$ creates a collinear quark moving along the $n$ direction, or
annihilates an antiquark.  The $b_v$ field is the standard usoft HQET field with
Lagrangian ${\cal L}_h=\bar b_v iv\mcdot D b_v$.  For a complete basis we also
need operators with octet bilinears. We take these to be $Q_i^{(0)}$ with
$T^A\otimes T^A$ color structure, for example
\begin{eqnarray}  
 Q_{\overline{1d}}^{(0)} = \big[ \bar u_{n,\omega_1} \bnslash P_L T^A b_v\big] 
  \big[ \bar d_{\bn,\omega_2} \nslash P_L T^A u_{\bn,\omega_3}
\big] \,.
\end{eqnarray} 
These $\overline {id}$ and $\overline {is}$ operators do not contribute to the
decays $B \to M_1 M_2$ at leading order, but will in power corrections.  Our
basis of ${Q}^{(0)}_{id}$ operators can be directly related to the one
derived in~\cite{chay}, except that we also included ${Q}^{(0)}_{3d}$ which
makes the basis sufficient to accommodate all electroweak penguin effects.

We also need the ${\cal O}(\lambda)$ operators for the LO factorization.
Defining 
\begin{eqnarray}
 ig\,{\cal B}^{\perp\,\mu}_{n,\omega} = \frac{1}{(-\omega)}\, 
 \big[ W^\dagger_n [ i\bn\mcdot D_{c,n} , i D^\mu_{n,\perp} ] W_n 
  \delta(\omega-\bnP^\dagger) \big]
\end{eqnarray} 
they are:
\begin{eqnarray}\label{Q1a1b}
  Q_{1d}^{(1)} \!\!&=&\!\! \frac{-2}{m_b} 
     \big[ \bar u_{n,\omega_1}\, ig\,\slash\!\!\!\!{\cal B}^\perp_{n,\omega_4} 
     P_L b_v\big]
     \big[ \bar d_{\bn,\omega_2}  \nslash P_L u_{\bn,\omega_3} \big] 
     \,, \\
  Q_{2d,3d}^{(1)} &=&  \frac{-2}{m_b}  
     \big[ \bar d_{n,\omega_1} \, ig\,\slash\!\!\!\!{\cal B}^\perp_{n,\omega_4} 
     P_L b_v \big]
     \big[ \bar u_{\bn,\omega_2} \nslash P_{L,R} u_{\bn,\omega_3} \big]
      \,,\nn \\
  Q_{4d}^{(1)} &=&  \frac{-2}{m_b} 
     \big[ \bar q_{n,\omega_1} \, ig\,\slash\!\!\!\!{\cal B}^\perp_{n,\omega_4} 
     P_L b_v \big]
     \big[ \bar d_{\bn,\omega_2} \nslash P_{L}\, q_{\bn,\omega_3} \big]
      \,,\nn \\
  Q_{5d,6d}^{(1)} &=& \frac{-2}{m_b} 
    \big[ \bar d_{n,\omega_1} \, ig\,\slash\!\!\!\!{\cal B}^\perp_{n,\omega_4} 
     P_L b_v \big]
    \big[ \bar q_{\bn,\omega_2} \nslash P_{L,R} q_{\bn,\omega_3} \big]
      \,, \nn\\
  Q_{7d}^{(1)} \!\!&=&\!\! \frac{-2}{m_b} 
   \big[ \bar u_{n,\omega_1}\, ig\,{\cal B}^{\perp\, \mu}_{n,\omega_4} 
    P_L b_v\big]
   \big[ \bar d_{\bn,\omega_2}  \nslash \gamma^\perp_\mu P_R u_{\bn,\omega_3} \big] 
      \,,\nn\\
  Q_{8d}^{(1)} \!\!&=&\!\! \frac{-2}{m_b} 
   \big[ \bar q_{n,\omega_1}\, ig\,{\cal B}^{\perp\, \mu}_{n,\omega_4} 
    P_L b_v\big]
   \big[ \bar d_{\bn,\omega_2}  \nslash \gamma^\perp_\mu P_R q_{\bn,\omega_3} \big] 
      \,. \nn
\end{eqnarray}
Our basis in Eq.(\ref{Q1a1b}) is simpler than the one in \cite{chay} for several
reasons. Terms with a ${\cal B}_n^\perp$ or $D_n^{\perp}$ in the $\bn$-bilinear
can be reduced to Eq.(\ref{Q1a1b}) by Fierz transformations.  This shows that
hard-spectator and form factor contributions are related. Second,
$\cPslash_\perp Q_{if}^{(0)}=0$, so integration by parts allows a basis for
$Q_{if}^{(1)}$ with no $n$-covariant derivatives, so only field strengths ${\cal
  B}_n^\perp$ appear, plus $[\bar u_n \gamma_\perp^\mu P_L b_v] {\cal
  P}^\mu_\perp [\bar d_\bn \nslash P_L u_\bn]$ terms which give vanishing
contributions.  We suppress $Q^{(1)}$'s with octet bilinears that do not
contribute at LO. The operators $Q_{5,6}^{(0,1)}$ only contribute to $SU(3)_\bn$
singlet production and are not used below.

\begin{table*}[t!]
%{@{*}r||p{1in}@{*}}[t!]
%\begin{center}
\begin{tabular}{|c|c|c||c|c|c|}
\hline
$M_1 M_2$ & $T_{1\zeta}(u)$ &  $T_{2\zeta}(u)$
& $M_1 M_2$ & $T_{1\zeta}(u)$ &  $T_{2\zeta}(u)$
\\
\hline\hline 
%%%%%%%%%%%%%%%%%%%%%%%%%%%%%%%%%%%%%%%%%%%%%%%%%%%%%%%%%%%%
$\pi^- \pi^+$, $\rho^-\pi^+$, $\pi^-\rho^+$, 
 $\rho^-_\parallel \rho^+_\parallel$  & 
  $c_1^{(d)}+c_4^{(d)}$ & 
  $0$ &
$\pi^+ K^{(*)-}$, $\rho^+ K^-$, $\rho_\parallel^+ K^{*-}_\parallel$  & 
  $0$ &
  $c_1^{(s)}+c_4^{(s)}$ 
\\
%%%%%%%%%%%%%%%%%%%%%%%%%%%%%%%%%%%%%%%%%%%%%%%%%%%%%%%%%%%%
$\pi^- \pi^0$, $\rho^-\pi^0$ & 
  $\frac{1}{\sqrt{2}}(c_1^{(d)}\!+\!c_4^{(d)}) $ &
  $\frac{1}{\sqrt{2}}(c_2^{(d)}\!-\!c_3^{(d)}\!-\!c_4^{(d)})$ &
$\pi^0 K^{(*)-}$ & 
  $\frac{1}{\sqrt{2}}(c_2^{(s)}\!-\!c_3^{(s)})$ &
  $\frac{1}{\sqrt{2}}(c_1^{(s)}\!+\!c_4^{(s)})$ 
\\
%%%%%%%%%%%%%%%%%%%%%%%%%%%%%%%%%%%%%%%%%%%%%%%%%%%%%%%%%%%%
$\pi^- \rho^0$, $\rho^-_\parallel \rho^0_\parallel$ & 
  $\frac{1}{\sqrt{2}}(c_1^{(d)}\!+\!c_4^{(d)})$ &
  $\frac{1}{\sqrt{2}}(c_2^{(d)}\!+\!c_3^{(d)}\!-\!c_4^{(d)})$ &
$\rho^0 K^-$, $\rho_\parallel^0 K^{*-}_\parallel$ &
  $\frac{1}{\sqrt{2}}(c_2^{(s)}\!+\!c_3^{(s)})$ &
  $\frac{1}{\sqrt{2}}(c_1^{(s)}\!+\!c_4^{(s)})$ 
\\
%%%%%%%%%%%%%%%%%%%%%%%%%%%%%%%%%%%%%%%%%%%%%%%%%%%%%%%%%%%%
$\pi^0 \pi^0$ &  
  $\frac{1}{2}(c_2^{(d)}\!-\!c_3^{(d)}\!-\!c_4^{(d)})$ &
  $\frac{1}{2}(c_2^{(d)}\!-\!c_3^{(d)}\!-\!c_4^{(d)})$ &
$\pi^- \bar K^{(*)0}$, $\rho^-\bar K^0$, 
 $\rho^-_\parallel \bar K^{*0}_\parallel$ & 
  $0$ &
  $-c_4^{(s)}$ 
\\

%%%%%%%%%%%%%%%%%%%%%%%%%%%%%%%%%%%%%%%%%%%%%%%%%%%%%%%%%%%%
$\rho^0 \pi^0$ &  
  $\frac{1}{2}(c_2^{(d)}\!+\!c_3^{(d)}\!-\!c_4^{(d)})$ &
  $\frac{1}{2}(c_2^{(d)}\!-\!c_3^{(d)}\!-\!c_4^{(d)})$ &
$\pi^0 \bar K^{(*)0}$ &  
  $\frac{1}{\sqrt{2}}(c_2^{(s)}\!-\!c_3^{(s)}) $ &
  $-\frac{1}{\sqrt{2}} c_4^{(s)}$ 
\\
%%%%%%%%%%%%%%%%%%%%%%%%%%%%%%%%%%%%%%%%%%%%%%%%%%%%%%%%%%%%
$\rho^0_\parallel \rho^0_\parallel$  & 
   $\frac{1}{2}(c_2^{(d)}\!+\!c_3^{(d)}\!-\!c_4^{(d)})$ &
  $\frac{1}{2}(c_2^{(d)}\!+\!c_3^{(d)}\!-\!c_4^{(d)})$ &
$\rho^0 \bar K^0$, $\rho^0_\parallel \bar K^{*0}_\parallel$ & 
  $\frac{1}{\sqrt{2}}(c_2^{(s)}\!+\!c_3^{(s)}) $ &
  $-\frac{1}{\sqrt{2}} c_4^{(s)}$ 
\\
%%%%%%%%%%%%%%%%%%%%%%%%%%%%%%%%%%%%%%%%%%%%%%%%%%%%%%%%%%%%
$K^{(*)0} K^{(*)-} $, $K^{(*)0} \bar K^{(*)0}$  & 
   $-c_4^{(d)}$ &
   $0$ &
  $K^{(*)-} K^{(*)+}$ & 
   $0$ &
  $0$ 
\\
  \hline\hline
\end{tabular}
%\end{center}
\vskip-4pt
{
\caption{
Combinations of Wilson coefficients appearing in the factorization formula. Note
that these results do not assume isospin symmetry and  all $VV$ channels in this 
table are longitudinal. Due to our basis choice the coefficients 
$T_{1\!J,2\!J}(u,z)$ for all these states are {\em identical} to 
$T_{1\zeta,2\zeta}(u)$ with each $c_i^{(f)}(u)$ replaced by $b_i^{(f)}(u,z)$.
}
\label{tab1} }
\vskip-13pt
\end{table*}

Next we determine the most general structure of the $p^2\sim E\Lambda$
contributions in \SCETa\!. We decouple the usoft modes by making the field
redefinitions~\cite{SCET} $\xi_{n'} \to Y_{n'} \xi_{n'}$, $A_{n'} \to Y_{n'}
A_{n'} Y_{n'}^\dagger$, with $Y_{n'}$ a Wilson line of $n'\mcdot A_{us}$ gluons
and $n'=n$ or $\bn$. In $Q_{if}^{(0,1)}$ all $Y$'s cancel except for
$(Y_n^\dagger b_v)$~\cite{chay}, and the operators factor into $(n,v)$ and $\bn$
parts,
\begin{eqnarray} \label{split}
 Q_{if}^{(0,1)} = \tilde Q_{if}^{(0,1)} Q_{if}^\bn \,.
\end{eqnarray}
In Fig.~\ref{fig:factor} the $M'$ meson only connects to the rest of the diagram
at the scale $p^2\sim Q^2$, through $Q_{if}^\bn=\bar q_{\bn,\omega_2} \Gamma
q'_{\bn,\omega_3}$ for some flavors $q,q'$ and Dirac structure $\Gamma$. The
shaded $p^2\sim E\Lambda$ region is required to generate the collinear $M$,
similar to the $B\to M$ form factors~\cite{bps4}.  At LO it is given by
$T$-products of the remaining parts of the operators in Eq.~(\ref{split}),
$\tilde Q_{if}^{(0,1)}$, with one Lagrangian ${\cal L}_{q\xi}^{(j)}$ inserted on
the spectator quark to swap it from usoft to collinear:
\begin{eqnarray} \label{Tproducts}
  T_1 \!\!&=&\!\! \mbox{\large $\int$} d^4y\, d^4y'\,
    T \big[\tilde Q^{(0)}_i(0) ,i{\cal L}^{(1)}_{\xi_n q}(y),i{\cal
    L}_{\xi_n\xi_n}^{(1)}\!(y')\! \nn \\
  &&\hspace{-0.1cm}
    + i{\cal L}_{cg}^{(1)}(y') \big] +  \mbox{\large $\int$} d^4y\, 
    T \big[\tilde Q_i^{(0)}(0),i{\cal L}^{(1,2)}_{\xi_n q}(y) \big], \ \nn\\
  T_2 \!\!&=&\!\! \mbox{\large $\int$} d^4y \:
    T \big[\tilde Q_i^{(1)}(0),i{\cal L}^{(1)}_{\xi_n q}(y) \big] .
\end{eqnarray}
%Analogous $T_1^{c\bar c}$ and $T_2^{c\bar c}$ time ordered products are generated by
%matching ${\cal Q}_{c\bar c}$ onto operators with zero and one collinear gluon.
Here ${\cal L}_{\xi_n q}^{(1)}= \bar q_{us} Y ig\,\slash\!\!\!\!{\cal B}^\perp_{n}
W^\dagger \xi_n +{\rm h.c.}$~\cite{bcdf}, and the form of our other ${\cal L}$'s
can be found in~\cite{bps5}.  

Now we match $\SCETa$ onto $\SCETb$\!.  A complete treatment of $T_1$ is an open
question due to endpoint singularities~\cite{bps4,LN,BFfinite}, but $\langle V_\perp |
T_1 | B\rangle =0$ and the nonzero matrix elements can be parameterized as
\begin{eqnarray} 
  \langle P | T_1 | B \rangle =  m_B\, \zeta^{BP}\,,\quad
  \langle V_\parallel | T_1 | B \rangle =  m_B\, \zeta^{BV_\parallel} \,.
\end{eqnarray}
For $T_2$ the most general perturbative matching at $\mu^2\sim E\Lambda$
generates a set of operators with Wilson coefficients given by jet functions $J$
and $J_\perp$ whose form is constrained by RPI, chirality, power counting and
dimensional analysis [\,$\omega_1 = z\omega$, $\omega_4=(1\!-\!z)\omega$, $\bar
x=1\!-\!x$\,, $\chi_{n,\omega}=(W^\dagger\xi_n)_{\omega}$],
\begin{eqnarray}\label{Jdef}
&& \hspace{-.4cm}
 T\,\big[ (\bar \xi_n W)_{\omega_1}
   i g\, {\cal B}^{\perp\alpha}_{n,\omega_4}
   P_{R,L} \big]^{ia}(0)\:
   \big[ i g\, \slash\!\!\!\!{\cal B}^\perp_{n}
   W^\dagger \xi_n \big]_0^{jb}(y) \nn
   \\
&=&\!\!\!
   i\, \delta^{ab} {\delta(y^+) \delta^{(2)}(y_\perp)}\,
    \frac{1}{\omega}
   \int_0^1\!\!\! dx \int\! \frac{dk^+}{2\pi}\: e^{+i  k^+ y^-/2}
   \nn\\
&\times&\!\!\!\! \Big\{\! -\!J_\perp(z,x,k_+)   \Big( \frac{\nslash}{2} P_{R,L}
  \gamma_\perp^\alpha \gamma_\perp^\beta \Big)_{ji}
  \big[\bar\chi_{n,x\omega}
   {\bnslash\gamma^\perp_\beta }  \chi_{n,-\bar x\omega}
   \big] \nn\\
&+&\!\!\!
  J(z,x,k_+) \Big({\nslash} P_{L,R} \gamma_\perp^\alpha\Big)_{ji}
  [\bar\chi_{n,x\omega} {\bnslash P_{L,R}} \chi_{n,-\bar
    x\omega}] \Big\},
\end{eqnarray}
where $\{i,j\}$ and $\{a,b\}$ are spin and color indices.  At tree level
we find that $J(z,x,k_+) =J_\perp(z,x,k_+) = {\delta(x-z) \pi \alpha_s(\mu)
  C_F}/{(N_c\, \bar x k_+)}$. The remaining pieces of $T_2$ are purely usoft and
match directly onto soft operators in \SCETb, giving
%\begin{mathletters}
\begin{eqnarray} \label{softpart}
 && -\frac{2 i}{m_b} \int\! d^4y\   
   [\, \bar q_{s} Y]^j(y)\   [  b_v]^{i}(0) 
   \label{Sdef} \,,\\
 && -\frac{2 i}{m_b} \int\! d^4y\   
   [\, \bar q_{s} Y]^j(y)\   [ \gamma^\perp_\alpha b_v]^{i}(0) \,, \label{Sdef2}
\end{eqnarray}
%\end{mathletters}
where here Eq.~(\ref{Sdef}) goes along with the $J_\perp$ term, and
Eq.~(\ref{Sdef2}) goes along with the $J$ term.

To obtain the final result for amplitudes we combine
Eqs.~(\ref{split}-\ref{softpart}), simplify the Dirac structure between the soft
fields, and take matrix elements. First consider final states containing
perpendicularly polarized vector mesons, $B\to V_\perp V_\perp$.
Kagan~\cite{alex} has argued that $B\to V_\perp V_\perp$ is power suppressed
relative to the longitudinal polarization, $B\to V_\parallel V_\parallel$.  At
LO in SCET $Q_{if}^{(0,1)}$ for $i=1$--$6$ have scalar bilinears and give
vanishing contributions to $B\to V_\perp V_\perp$.  The operators $\tilde
Q_{7f}^{(1)}$ and $\tilde Q_{8f}^{(1)}$ generate the $J_\perp$ term in
Eq.~(\ref{Jdef}) and could contribute. However, chirality conservation in \SCETa
implies that one vector is L and one is R polarized so the $\tilde Q_{7f}^{(1)}$
and $\tilde Q_{8f}^{(1)}$ contributions also vanish (quark masses flip chirality
and in \SCETa are suppressed by powers of $m_q/\sqrt{\Lambda E}$~\cite{wise}).
More explicitly the $J_\perp$ term in Eq.~(\ref{Jdef}) vanishes because the soft
Dirac structure can be reduced, $\nslash P_L
\gamma_\perp^\alpha\gamma_\perp^\beta =
(g_\perp^{\alpha\beta}+i\epsilon_\perp^{\alpha\beta})\, \nslash P_L$, and this
tensor vanishes when contracted with the $\bn$-bilinear,
\begin{eqnarray}
 (g_\perp^{\alpha\beta}+i\epsilon_\perp^{\alpha\beta}) \: 
  \bar d_{\bn,\omega_2}\nslash \gamma^\perp_\alpha P_R\, q_{\bn,\omega_3}=0 \,.
\end{eqnarray}
Thus at LO only $A_{c\bar c}$ could give transverse polarized vector mesons so 
\begin{eqnarray}
 A(B\to V_1^\perp V_2^\perp) = 
  \frac{2G_F}{\sqrt{2}}\:
  \langle V_1^\perp V_2^\perp |\: Q_{c\bar c}  | B \rangle \,.
\end{eqnarray}

Next consider $B\to V_\parallel V_\parallel$, $B\to V_\parallel P$ and $B\to PP$
decays. Now it is the $J$ term in Eq.~(\ref{Jdef}) that contributes along with
possible long distance charming penguins.  Due to the form of our operators the
$J$ term is identical to the analysis of the $B\to M$ form factors.  The LO
factorization formula for $A=\langle M_1 M_2 | H_W |B\rangle$ which determines
$\bar B^0, B^- \to M_1 M_2$ with $M_{1,2}$ pseudoscalars or longitudinal vectors
is
\begin{eqnarray}\label{fact_general}
A(\bar B\to M_1 M_2) &=& 
\lambda_c^{(f)} A^{M_1M_2}_{c\bar c} +
\frac{G_F m_B^2}{\sqrt2}  \bigg\{ 
  f_{M_2}\, \zeta^{BM_1} \nn\\
 &&\hspace{-3cm}
  \times\! \int_0^1\!\!\!\! du\: T_{2\zeta}(u)\, \phi^{M_2}(u) + 
  f_{M_1}\, \zeta^{BM_2} \int_0^1\!\!\!\! du\: T_{1\zeta}(u)\, \phi^{M_1}(u)
  \nn \\
& & \hspace{-3cm}
  +\frac{f_B f_{M_1} f_{M_2}}{m_b} \int_0^1\!\!\!\!du \int_0^1\!\!\!\! dx 
  \int_0^1\!\!\!\! dz \! \int_0^\infty\!\!\!\!\!\! dk_+ \,  J(z,x,k_+)
  \Big[T_{2J}(u,z) \nn\\
 &&\hspace{-3cm}
  \times \phi^{M_1}(x) \phi^{M_2}(u) 
  + T_{1J}(u,z) 
  \phi^{M_2}(x) \phi^{M_1}(u)\Big]\phi_B^+(k_+) \bigg\}  \nn \,, \\
\end{eqnarray}
where $A^{c\bar c}$ denote possible long distance charming penguin amplitudes
which contribute in channels where $c_4^{(d,s)}$ appear.  For each decay mode
the set of hard coefficients $T_{i\zeta}$ and $T_{iJ}$ can be obtained from 
Table~\ref{tab1}. 

A new result from our analysis is that the jet function $J$ in
Eq.~(\ref{fact_general}) is the {\em same} as that appearing in the
factorization formula for $B\to M$ form factors~\cite{ps1}. We quote here two of
these formulas, one for the standard $B\to P\ell\bar\nu$ form factor $f_+(E)$,
and one for the form factor $A_\parallel$ for $B\to V_\parallel \ell\bar\nu$
decays,
\begin{eqnarray}
  A_\parallel(E) \!=\! \frac{1}{m_V}\Big[
    \frac{m_B E\, A_2(E) }{m_B\!+\!m_V} 
   \!-\! 
   \frac{(m_B\!+\!m_V)}{2}\, A_1(E)\Big]
   ,
\end{eqnarray}
where
\begin{eqnarray} 
  E=\frac{m_B^2+m_M^2-q^2}{2 m_B} \,.
\end{eqnarray} 
At LO in SCET~\cite{bf,bps4,ps1,BFfinite,LN}
\begin{eqnarray}\label{ff_factorization}
f_+(E) \!\!&=&\!\! T^{(+)}(E)\, \zeta^{BP}(E)
  + N_0 \int_0^1\!\!\!\! dz\! \int_0^1\!\!\!\! dx\! 
   \int_0^\infty\!\!\!\!\! dk_+ \,
    \nn \\
& &\hspace{-1cm}
 \times C_J^{(+)}(z,E) J(z,x,k_+,E) \phi^{M}(x) \phi_B^+(k_+)
   \,, 
 \nn \\[5pt]
%\end{eqnarray}
%\begin{eqnarray} \label{ff_factorization2}
%
 A_\parallel(E) \!\!&=&\!\! 
T^{(\!A_\parallel)}(E)  \zeta^{BV_\parallel}(E)  +  N_\parallel 
\int_0^1 \!\!\!\! dz\!\int_0^1\!\!\!\! dx \int_0^\infty \!\!\!\!  dk_+ 
\nonumber\\
& &\hspace{-1cm} \times C_J^{(\!A_\parallel)}(z,E)  J(z,x,k_+,E) 
  \phi^{M}(x) \phi_B^+(k_+) ,
\end{eqnarray}
where $N_0 = f_B f_P m_B/(4E^2), N_\parallel = f_B f_V m_B/(4E^2)$, and the
functions $T^{(+,A)}(E), C_J^{(+,A)}(z)$ are combinations of SCET Wilson
coefficients and can be found in~\cite{ps1}.  In that paper the jet functions
$J^{(\perp)}(z,x,k_+)$ in Eq.~(\ref{Jdef}) are denoted by
$J_{b}^{(\perp)}(z,x,k_+)$ and $J_a^{(\perp)}(x,k_+) = \int_0^1\! dz\,
J_b^{(\perp)}(x,z,k_+)$.  At the endpoint where $E\simeq m_B/2$ the same
parameters $\zeta^{BM}$ and jet function $J$ appear in the form factors and in
the non-leptonic decays.  Since the analysis for $J$ is identical to that in the
form factors several important facts can be immediately taken over for $B\to M_1
M_2$ decays.  In particular to all orders in perturbation theory only the
$\phi_B^+(k_+)$ wavefunction is obtained as proven in Ref.~\cite{ps1}.  Also the
convolution integrals with $J$ are finite with an identical proof to the one
given in Ref.~\cite{BFfinite}.  Finally it is clear that possible messenger
fluctuations~\cite{messenger} can not spoil factorization in ${\cal
  Q}^{(0,1)}_{if}$ which have color singlet $\bn$-bilinears, and so their role
will be identical to that in the form factors.

At this point we compare our result in Eq.~(\ref{fact_general}) with the result in QCDF~\cite{BBNS}.  From Eq.~(25) of~\cite{BBNS} the LO factorization theorem is
\begin{eqnarray} \label{factbbns}
 \langle M_1 M_2 | O_i | B\rangle &=&   \\
  && \hspace{-2.4cm}
 \Big\{ F^{B\to M_1}(0) f_{M_2} \int\!\! du\, T^I_{M_2,i}(u)
  \phi_{M_2}(u) + {(1 \leftrightarrow 2)} \Big\}
  \nn\\
 &&\hspace{-2.4cm}
  + f_{M_1} f_{M_2} f_{B} \int\!\! du\, dx\, dk^+ \nn \\
  &&\hspace{-2cm}
  \times T^{\rm II}_i(x,u,k_+)  \phi_{M_1}(x) \phi_{M_2}(u) \phi_B(k_+) \,,
\end{eqnarray}
where the parameters are the QCD form factors $F^{B\to M}(0)$, $\phi_{M_i}$, and
$\phi_B$ (other parameters appear when power suppressed terms from annihilation
or chirally enhanced corrections are included). In the QCDF power counting the
second term is suppressed relative to the first by a factor of $\alpha_s$.  The
result in Eq.~(\ref{factbbns}) is quite similar to the SCET formula derived in
Eq. (\ref{fact_general}). However, there are several important differences,
which we comment on.  The two things that are most important for phenomenology
are that QCDF does not allow for a leading order $A_{c\bar c}^{\pi\pi}$
contribution, and that the SCET analysis suggests that the contribution from
$\zeta$ and $\zeta_J$ are comparable in size, rather than $\zeta_J^{B\pi}\ll
\zeta^{B\pi}$ as in QCDF.  As discussed later, current data on $B \to \pi \pi$
seems to support $\zeta_J^{B\pi} \sim \zeta^{B\pi}$, albeit with large
uncertainties.  This difference has significant phenomenological ramifications,
as it implies that even in absence of leading order charming penguin effects the
perturbative strong phases predicted in~\cite{BBNS} would receive ${\cal
  O}(100\%)$ corrections. Besides these points there are several technical
differences between the two formulas.  Using $F^{B\to M}(0)$ in
Eq.~(\ref{factbbns}) rather than $\zeta^{BM}$ does not completely separate out
all contributions from the hard scale. Also, in Eq.~(\ref{factbbns}) $T^{\rm I}$
and $T^{\rm II}$ include perturbative contributions from both the $\mu^2\simeq
Q^2$ and $\mu^2\simeq E\Lambda$ scales~\cite{bf}.  In the result in
Eq.~(\ref{fact_general}) these scales are separated in $T_{iJ}$ and $J$
respectively.  If $\zeta^{BM}$ is independent of the $\mu^2\simeq E\Lambda$
scale as argued in Ref.~\cite{LN} then the scales are also completely separated
in the $T_{i\zeta}\,\zeta^{BM}$ term, otherwise $\zeta^{BM}$ still encodes
physics at both the jet scale $E\Lambda$ and the scale $\Lambda^2$.

The jet function $J$ depends on physics at the intermediate scale, so its
perturbative expansion in $\alpha_s(\sqrt{E\Lambda})$ is not as convergent as
for the $T_{iJ}$ and $T_{i\zeta}$ which are expanded in $\alpha_s(Q)$. In fact
perturbation theory may fail for $J$ all together. This can be tested both by
experiment~\cite{mps} and by additional perturbative calculations. Using SCET we
can still obtain an expression for $A(\bar B\to M_1 M_2)$ without expanding
$J$ perturbatively,
\begin{eqnarray}  \label{A0newfact}
A \!\!&=&\!\! 
%A_{0}^{c\bar c}\!+\!
\frac{G_F m_B^2}{\sqrt2}\!\bigg\{
   f_{M_1}\! \int_0^1\!\!\!\!du\, dz\,
    T_{1\!J}(u,z) \zeta^{BM_2}_{J}(z) \phi^{M_1}(u) 
   \\
 &&\hspace{-0.7cm}
   + f_{M_1} \zeta^{BM_2}\!\! \int_0^1\!\!\!\! du\, T_{1\zeta}(u) \phi^{M_1}(u)
  \bigg\} \!+\! \Big\{ 1\leftrightarrow 2\Big\} 
 \!+\! \lambda_c^{(f)} A_{c\bar c}^{M_1M_2}, \nn 
 % \nn\\
 %&&\hspace{-0.2cm} 
 %  + \lambda_c^{(f)} A_{c\bar c}^{M_1M_2}   \,,
\end{eqnarray}
where power counting implies $\zeta^{BM}\sim \zeta^{BM}_J\sim
(\Lambda/Q)^{3/2}$.  Here the non-perturbative parameters $\zeta^{BM}$,
$\zeta^{BM}_{J}(z)$, and $\phi^M(u)$, still all occur in the $B\to M$
semileptonic and rare form factors. For a model independent analysis they need
to be determined from data.  Note that it was possible for us to derive
Eq.~(\ref{A0newfact}) because in Eq.~(\ref{fact_general}) we separated the
scales $Q^2$ and $E\Lambda$ into $T$'s and $J$'s respectively.  The
corresponding results for the form factors in Eq.~(\ref{ff_factorization})
are
\begin{eqnarray}  \label{ffnewfact}
f_+\!\!&=&\!\! T^{(+)}(E)\,\zeta^{BP}\!(E) + N_0\!
\int_0^1\!\!\! dz \, C_J^{(+)}(z)\, \zeta_J^{BM}(z,E)  \,,\nn
 \\
A_\parallel \!\!&=&\!\! 
T^{(\!A_\parallel)}(E) \, \zeta^{BV_{\!\parallel}}\!(E) +  N_\parallel 
\int_0^1\!\!\! dz \, 
 C_J^{(\!A_\parallel)}(z)\, \zeta_J^{BV}(z,E) \nonumber \,. \\[-5pt]
\end{eqnarray}
The two form factors in Eq.~(\ref{ff_factorization}) can be obtained from data
on $B\to (P,V_\parallel)\ell \nu$, giving important information on the
$\zeta^{BM}, \zeta_J^{BM}$ appearing in Eq.~(\ref{fact_general}). Note that in
Eqs.~(\ref{fact_general}) and (\ref{A0newfact}) the $\zeta$'s are evaluated at
$E=m_B/2$.  Eq.~(\ref{fact_general}) and (\ref{A0newfact}) are the main results
of our paper.

Using Eq.~(\ref{A0newfact}) still requires matching the full
theory $O_i$'s onto the $Q_{if}^{(0,1)}$ to determine the Wilson coefficients
$c_i^{(f)}$ and $b_i^{(f)}$.  For the coefficients of $Q_i^{(0)}$ we find [$f=d,s$]
\begin{eqnarray}
  c_1^{(f)} \!\!&=&\!\! 
       \lambda_u^{(f)}\Big( C_1 \!+\! \frac{C_2}{N_c} \Big)
       -\lambda_t^{(f)} \frac32 \Big( C_{10} \!+\! \frac{C_9}{N_c}\Big)
       + \Delta c_1^{(f)}
      \,,\nn\\
  c_2^{(f)} \!\!&=&\!\! 
       \lambda_u^{(f)}\Big( C_2 \!+\! \frac{C_1}{N_c}\Big) 
       -\lambda_t^{(f)} \frac32 \Big(C_9 \!+\! \frac{C_{10}}{N_c}\Big)
       +  \Delta c_2^{(f)}
      \,,\nn \\
  c_3^{(f)} \!\!&=&\!\!
       -\lambda_t^{(f)}\frac32  \Big( C_7 + \frac{C_8}{N_c}\Big)
       +  \Delta c_3^{(f)}
     \,, \nn \\
  c_4^{(f)} \!\!&=&\!\! 
      -\lambda_t^{(f)}\Big( C_4 + \frac{C_3}{N_c}
       -\frac{C_{10}}{2} - \frac{C_9}{2N_c} \Big)
      +  \Delta c_4^{(f)}
     \,.
 %\\
%  c_5^{(d,s)} \!\!&=&\!\!
%      -\lambda_t^{(d,s)}\Big( C_3 + \frac{C_4}{N_c} - \frac{C_7}{2} 
%     - \frac{C_8}{2N_c}  \Big) 
%     +  \Delta c_5^{(d,s)}(\omega_j)
%     \,,\nn\\
%  c_6^{(d,s)} \!\!&=&\!\!
%     -\lambda_t^{(d,s)}\Big( C_5 + \frac{C_6}{N_c} 
%     - \frac{C_9}{2} - \frac{C_{10}}{2N_c} \Big) 
%     +  \Delta c_6^{(d,s)}(\omega_j)
%     \,,\nn
\end{eqnarray}
and for the $Q_i^{(1)}$ 
\begin{eqnarray}
   b_1^{(f)}  \!\!&=&\!\!
    \lambda_u^{(f)}\Big[ C_1 + \Big(1 \!-\!\frac{m_b}{\omega_3} \Big)
      \frac{C_2}{N_c} \Big] 
      \\
&& 
    - \lambda_t^{(f)}\Big[ \frac{3}{2} C_{10} + 
    \Big(1 \!-\!\frac{m_b}{\omega_3} \Big)
      \frac{3 C_9}{2 N_c} \Big]
     + \Delta b_1^{(f)}
     \,,\nn
     \\
   b_2^{(f)}  \!\!&=&\!\!
     \lambda_u^{(f)}\Big[ C_2 + \Big(1 \!- \! \frac{m_b}{\omega_3}  \Big)
      \frac{C_1}{N_c}\Big]
      \nn\\
     && 
     - \lambda_t^{(f)}\Big[ \frac{3}{2} C_9 + \Big(1 \!
     -\!\frac{m_b}{\omega_3} \Big)
      \frac{3 C_{10}}{2 N_c} \Big]
     + \Delta b_2^{(f)}
     \,,\nn 
     \\
   b_3^{(f)}  \!\!&=&\!\!
     - \lambda_t^{(f)}\Big[ \frac{3}{2} C_7 
     + \Big(1 \!-\!\frac{m_b}{\omega_2} \Big)
      \frac{3 C_8}{2 N_c} \Big]
     + \Delta b_3^{(f)}
     \,, \nn \\
   b_4^{(f)}  \!\!&=&\!\!
     -\lambda_t^{(f)}\Big[ C_4 \!-\! \frac{C_{10}}{2}\!
     +\!  \Big(1 \!-\! \frac{m_b}{\omega_3}  \Big)
      \Big( \frac{C_3}{N_c} \!-\! \frac{C_9}{2N_c} \Big)\Big]
      + \Delta b_4^{(f)} ,
\nn 
%\\
%&& 
%          \nn\\
%
\end{eqnarray}
%\begin{eqnarray}
%  b_7^{(f)}  &=&       
%     -\lambda_t^{(d,s)} \Big(\frac{m_b}{\omega_2}\!- \!\frac{m_b}{\omega_3}\Big)
%      \frac{3 C_9}{2N_c} 
%      \!+\!  \Delta b_7^{(d,s)}
%      \,,\nn\\
%  b_8^{(f)}  &=&
%      -\lambda_t^{(d,s)} \Big(\frac{2m_b}{\omega_2}\!
%      -\!\frac{2m_b}{\omega_3}\Big)
%      \Big(\frac{C_5}{N_c}\!-\!\frac{C_9}{2N_c}\Big)
%       \!+\!  \Delta b_8^{(d,s)}
%       \,. \nn
%\end{eqnarray}
where $\omega_2 = m_b u, \omega_3=-m_b \bar u = m_b(u-1)$ and the $\Delta
c_i^{(f)}$ and $\Delta b_i^{(f)}$ are perturbative corrections. The ${\cal
  O}(\alpha_s)$ contribution to the $\Delta c_j^{(f)}(u)$ have been calculated
in~\cite{BBNS} and later in~\cite{chay}. It is possible that these results will
need to be modified by an additional subtraction for the long distance charming
penguin. Finally, any full $\alpha_s(m_b)$ analysis requires $\Delta
b_j^{(f)}(u,z)$ which are currently unknown, unless the numerical values of
$\zeta$, $\zeta_J$ are such that $\zeta_J\sim \alpha_s(m_b)\zeta$ so that
$\zeta_J^{BM}\ll \zeta^{BM}$ and the $\Delta c_j^{(f)}$ coefficients dominate
numerically.

There are several issues in the phenomenological use of the factorization
formula. There is a hierarchy due to CKM factors and the $C_i$'s which have to
be accounted for in the $c_i^{(f)}$ and $b_i^{(f)}$. For example, $C_1$ is about
a factor of six larger than any of the other coefficients, making $c_1^{(d)}$,
$b_1^{(d)}$, and $b_2^{(d)}$ large. We will refer to quantities as
``contaminated'' if $1/m_b$ power corrections could compete with LO results due
to the heirarchy in Wilson coefficients.  Unless these corrections can be
accounted for or proven to be absent, one should assign $\sim 100\%$ uncertainty
to predictions for contaminated decays. The determination of whether a quantity
is contaminated depends on the relative size of $\zeta^{BM}$ and $\zeta_J^{BM}$.
If $\zeta^{BM}\gg \zeta_J^{BM}$ as in QCDF then any $f=d$ decay in
Table~\ref{tab1} that is independent of $c_1^{(d)}$ could receive large
corrections, making quantities such as $Br(\bar B^0 \to \pi^0\pi^0)$
contaminated~\cite{BBNS2}.  Here the most problematic are large power
corrections proportional to $C_1 \Lambda/E$ which is $\sim C_2$ and $\gg C_{i\ge
  3}$.  These can arise for example from T-products involving the $Q_{\overline
  {2f}}^{(0)}$ operators.  The situation is much better in the case
$\zeta^{BM}\sim \zeta_J^{BM}$ since any decay depending on $c_1^{(d)}$,
$b_1^{(d)}$, or $b_2^{(d)}$ will not be contaminated and can be expected to have
power corrections of normal size, $\sim 20\%$. Our analysis of $B\to \pi\pi$
below favors this situation, in which case $Br(\bar B^0 \to \pi^0\pi^0)$ is not
contaminated.
% In this case
%there still could be important perturbative corrections $\propto \alpha_s(m_b)
%C_1$, which have not yet been computed for the $b_i^{(f)}$'s.

At leading order in $\Lambda/E$ there are only two sources of strong phases: the
one-loop $\Delta c_i, \Delta b_i$ which can become complex \cite{BBNS}, and the
unfactorized $A_{c\bar c}$ charming penguin.  Additional final state phases come
from power corrections $\sim \Lambda/E$.  It is known from $\bar B^0\to
D^0\pi^0$ decays that $\Lambda/E$ corrections produce $\sim 30^\circ$
non-perturbative strong phases in agreement with dimensional
analysis~\cite{mps}. These large phases have nothing to do with a $\Lambda/m_c$
expansion so we expect strong phases of similar size from power corrections in
$B\to M_1 M_2$.  For contaminated decays, such as $B \to K K$, non-perturbative
strong phases $\propto C_1$ could be order unity.

The factorization theorems in Eqs.~(\ref{fact_general},\ref{A0newfact}) can be
used to make quantitative predictions for nonleptonic $B\to MM'$ decays.  There
are many applications; a few of the more important categories are: i)
Decay modes which are independent of charming penguin contributions are
determined by $\zeta$ and $\zeta_J$ which can be extracted from semileptonic
form factors.  ii) SCET implies SU(3) relations beyond those following
from $H_W$ in Eq.~(\ref{Hw}) with full QCD. It also simplifies the structure of
SU(3) breaking corrections.  iii) For $B\to VV'$ SCET allows us to
analyze polarization effects.  iv) Using isospin SCET makes predictions
for matrix elements whose quantum numbers differ from the reduced set of
$A_{c\bar c}^{M_1M_2}$ amplitudes. In the remainder of the paper we discuss
examples in each of these categories.  In particular we show that
Eq.~(\ref{A0newfact}) gives a reasonable fit to the current $B\to \pi\pi$ data.

The parameters $\zeta^{BM}$ and $\zeta_J^{BM}$ in Eq.~(\ref{A0newfact}) for
nonleptonic decays are common to those appearing in $B\to M$ form factors
Eq.~(\ref{ffnewfact}).  Decays that do not depend on $A_{c\bar c}$ include all
combinations in Table~\ref{tab1} that are independent of $c_4$ and $b_4$, such
as $B^- \to \pi^0\pi^-$ and $B^- \to \rho^0\rho^-$ once isospin is used.  For
example,
\begin{eqnarray}
&& \hspace{-1cm} \sqrt2 A(B^-\to \pi^-\pi^0) = 
\frac{G_F m_B^2}{\sqrt2} f_\pi \\
&&\hspace{-0.5cm}\times \bigg\{
  \int_0^1\!\!\!\!du\, dz\,
    (b_1^{(d)} \!+\! b_2^{(d)} \!-\! b_3^{(d)})(u,z) \,
    \zeta^{B\pi}_{J}(z) \phi^{\pi}\!(u) 
  \nn \\
  &&\hspace{-0.4cm}
   + \zeta^{B\pi} \int_0^1\!\!\!\! du\,  
(c_1^{(d)} \!+\! c_2^{(d)} \!-\! c_3^{(d)})(u)\, \phi^{\pi}\!(u)
  \bigg\}  \,, \nn
\end{eqnarray}
At tree level the $b_i^{(f)}$'s are independent of $z$ and this relation gives a
clean constraint on $\zeta^{B\pi}$ and $\zeta_J^{B\pi}=\int\! dz\,
\zeta_J^{B\pi}(z)$.

Flavor SU(3) symmetry is a powerful tool for studying nonleptonic B decays.  In
one particular application, Ref.~\cite{su3} proposed using flavor SU(3) symmetry
to determine $\gamma$ from $B^+\to K\pi, \pi^+\pi^0$.  Corrections to this
approach come from SU(3) breaking effects and are typically $\sim 30\%$.  The
factorization relation Eq.~(\ref{A0newfact}) implies enhanced SU(3) relations
beyond those in QCD. For example, in QCD all $B\to PP$ decays to two
pseudoscalar octet mesons are parameterized in the SU(3) limit by 5 complex
amplitudes. Using the SCET factorization formula Eq.~(\ref{fact_general}) this
number is reduced to one complex amplitude $A_{c\bar c}$, one real number
$\zeta$ and one real function $\zeta_J(z)$. In the language of Ref.~\cite{su3}
the operators in Eq.~(\ref{match}) do not generate the $E$, $A$, and $P\!A$
amplitudes, so these are power suppressed.

In certain cases the SU(3) breaking can be also computed. Such an example is the
determination of two SU(3) breaking parameters $R_{1,2}$ appearing in a SU(3)
relation used to extract $\gamma$~\cite{su3}
\begin{eqnarray}
& &\hspace{-0.5cm}
 A(B^- \to \bar K^0 \pi^-) + \sqrt2 A(B^- \to K^- \pi^0) =\\
& &\sqrt2 \, \frac{|V_{us}|}{|V_{ud}|} (R_1 - \delta_{EW} e^{i\gamma} R_2)
 A(B^- \to \pi^- \pi^0)\nonumber \,.
\end{eqnarray}
Here $\delta_{EW}$ parameterizes the largest electroweak penguin effects and is
calculable.  The parameters $R_{1,2}$ can be expressed in terms of
$\zeta^{B\pi},\zeta^{BK}, \zeta_J^{B\pi}(z), \zeta_J^{BK}(z)$ and calculable
Wilson coefficients and do not involve $A_{c\bar c}^{\pi\pi}$ or $A_{c\bar
  c}^{K\pi}$.

Polarization measurements in decays to two vector mesons have received much
attention recently. These decays were studied in Ref.~\cite{alex}, and it was
argued that factorization implies $R_T \sim 1/m_b^2$ and $R_\perp/R_\parallel =
1 + {\cal O}(1/m_b)$, where $R_{0,T,\perp,\parallel}$ denote the longitudinal,
transverse, perpendicular and parallel polarization fractions
($R_T=R_\perp+R_\parallel$, $R_0+R_T=1$). Using SCET we find that $R_T$ is power
suppressed in agreement with~\cite{alex}, unless the charming penguin amplitude
$A_{c\bar c}$ spoils this result.  We can not resolve the validity of the
$R_\perp/R_\parallel$ relation working only at LO in $1/m_b$.  Experimentally,
one finds \cite{rhoexp,phiKstexp}
\begin{eqnarray}
R_0(B^+\to \rho^+\rho^0) &=& 0.975 \pm 0.045 \,, \\
R_0(B^0\to \rho^+\rho^-) &=& 0.98^{+0.02}_{-0.08} \pm 0.03  \nonumber \,, \\
R_0(B^0\to \phi K^{*}) &=& 0.49 \pm 0.06 \,.\nonumber
\end{eqnarray}
It has been argued that the large transverse polarization observed in the $\phi
K^*$ mode might provide a second hint at new physics in $b\to s\bar s s$
channels beyond $\sin(2\beta)$ from $B\to \phi K_S$.  Unfortunately this
conclusion could be spoiled by a contribution from $A_{c \bar c}$ at leading
order.  $A_{c\bar c}$ does not contribute to $B^+\to \rho^+\rho^0$, but can
affect $B^0\to \phi K^{*}$ and $B^0\to \rho^+\rho^-$. Until charming penguins
are better understood the polarization measurements do not provide a clean
signal of physics beyond the standard model.

We finally examine in some detail the predictions of this paper for $B\to \pi\pi$
decays, and show that they reproduce the existing data.  The present world
averages are \cite{HFAG}
\begin{eqnarray}\label{pipidata}
& &S_{\pi\pi} = -0.74\pm 0.16,\quad  C_{\pi\pi} = -0.46\pm  0.13\,,\nn \\
& &Br(B^+\to \pi^0\pi^+) = (5.2\pm 0.8)\times 10^{-6} \,, \nn \\
& &Br(B^0\to \pi^+\pi^-) = (4.6\pm 0.4)\times 10^{-6} \,, \nn \\ 
& &Br(B^0\to \pi^0\pi^0) = (1.9\pm 0.5)\times 10^{-6} \,,
\end{eqnarray}
where the branching fractions are CP averages.  The amplitudes are naturally
divided into two pieces with different CKM factors, as $A \equiv \lambda_u^{(d)}
T + \lambda_c^{(d)} P$, where $T$ and $P$ are usually called ``tree'' and
``penguin'' amplitudes.  The decay amplitudes for $B \to \pi \pi$ can be written
in a model-independent way as
\begin{eqnarray}\label{b2pipi}
A(\bar B^0\to \pi^+\pi^-) &=& \lambda_u^{(d)}\, T_c 
(1 + r_c\, e^{i\delta_c} e^{i\gamma}) \,, \nn\\
\nonumber
A(\bar B^0\to \pi^0\pi^0) &=& \lambda_u^{(d)}\, T_n 
(1 + r_n\, e^{i\delta_n} e^{i\gamma}) \,,\\
\sqrt2 A(B^-\to \pi^0\pi^-) &=& \lambda_u^{(d)}\, T  \,,
\end{eqnarray}
where $(r_c,\delta_c)$ and $(r_n,\delta_n)$ parameterize the ratio of tree to
penguin contributions to $B^0\to \pi^+\pi^-$ and $B^0\to \pi^0\pi^0$,
respectively. We have neglected small electroweak penguin contributions.
Isospin gives the relations
\begin{eqnarray}\label{isospin}
  T = T_c + T_n\,,\quad
  T_c r_c e^{i \delta_c} + T_n r_n e^{i \delta_n} = 0 \,,
\end{eqnarray}
leaving only 5 independent strong interaction parameters in Eq.~(\ref{b2pipi}).

In the first step of the analysis, we assume that $\beta,\gamma$ are known, use
this to disentangle the tree and penguin amplitudes, and thus extract the five
parameters in Eq.~(\ref{b2pipi}).  In a second step, these parameters are
compared with the leading order predictions from SCET, and used to extract the
nonperturbative parameters appearing in the factorization formula
Eq.~(\ref{A0newfact}), working at tree level in matching at the hard scale. The
resulting SCET parameters are then used to predict values for $|V_{ub}| f_+(0)$
and $Br(B^0\to \pi^0\pi^0)$ as functions of $\gamma$.

\begin{figure}[t!]
  \centerline{ \mbox{\epsfysize=8truecm \hbox{\epsfbox{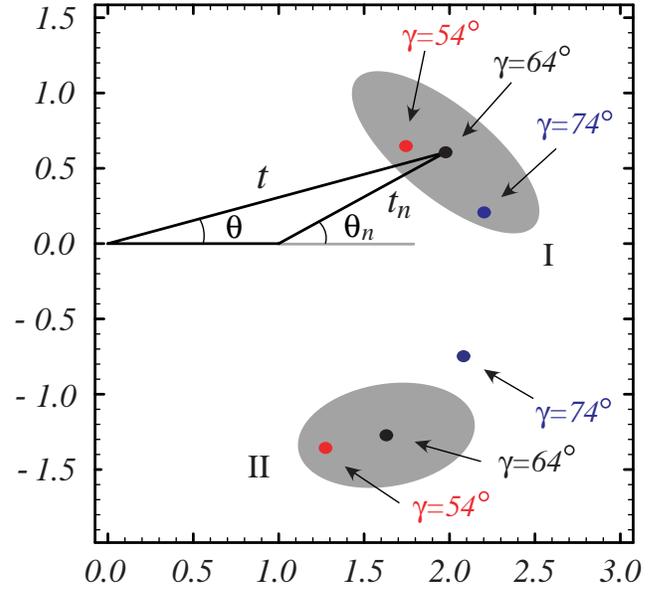}} } }
  \vskip-0.2cm {\caption[1]{ Constraints on the triangle of tree amplitudes
      $T/T_c - T_n/T_c = 1$ from current world averaged data on $B\to \pi\pi$.
      The shaded regions show the two $1$-$\sigma$ regions for $\gamma=64^\circ$
      including the error correlation between $|t|$ and $|t_n|$. The central
      values for $\gamma=54^\circ$ and $\gamma=74^\circ$ are also shown.}
\label{fig:pipidata} }
\vskip -0.4cm
\end{figure}
Assuming values for the CKM angles $\beta$ and $\gamma$ we can use the 5 pieces
of experimental data given in Eq.~(\ref{pipidata}) to determine the 5 parameters
in Eq.~(\ref{b2pipi}). Using $(\beta, \gamma) = (23^\circ,64^\circ)$ and the
data for the CP asymmetries we find for the penguin parameters $r_c$ and
$\delta_c$
\begin{eqnarray} \label{rcdelc}
r_c = 0.75\pm 0.35\,,\quad \delta_c = -44^\circ\pm 12^\circ\,.
\end{eqnarray}
This is in good agreement with the recent determinations of these parameters in
Refs.~\cite{pipidata}.  Using the branching ratio data as input, we can
determine the tree parameters as well. We find
\begin{eqnarray}\label{Texp}
 |T| &=&
  N_\pi\, ({0.29 \pm 0.02}) \Big( \frac{3.9\times 10^{-3}}{|V_{ub}|} \Big)\,,\\
 |t| &=& 2.07\pm 0.42\,,\quad 
 |t_n| =
\left\{
\begin{array}{cc}
1.15\pm 0.33 & \mbox{(I)}\\
1.42\pm 0.35 & \mbox{(II)}\\
\end{array}
\right.\nn \,,
\end{eqnarray}
where $N_\pi = \frac{G_F}{\sqrt2} m_B^2 f_\pi$ and we defined 
\begin{eqnarray}
t = \frac{T}{T_c}\,,\quad  \qquad t_n = \frac{T_n}{T_c} \,.
\end{eqnarray}
Some of the errors in Eqs.~(\ref{rcdelc}) and (\ref{Texp}) have sizeable
correlations.  The results for the tree triangle are shown graphically in
Fig.~\ref{fig:pipidata}.  The two $\gamma=64^\circ$ solutions correspond to
those in Eq.~(\ref{Texp}) and the ellipses denote $1\sigma$ contours. Also shown
in this figure is the isospin tree triangle, which for the reduced tree level
amplitudes reads $1 + t_n = t$.  There are two strong phases in this triangle
which are also shown in the figure, namely $\theta$ between $T$ and $T_c$ and
$\theta_n$ between $T_n$ and $T_c$.

As a second step the extracted amplitudes are compared with the predictions
of this paper at leading order in $\Lambda/m_b$ and tree level in the SCET
Wilson coefficient $c_i^{(d)}$ and $b_i^{(d)}$. At this order our result has
four independent parameters. The tree amplitudes $T,T_c$ are given by the
factorization relation Eq.~(\ref{fact_general}) and depend on the
non-perturbative parameters $\zeta^{B\pi},\zeta_J^{B\pi}$, 
\begin{eqnarray}\label{T}
T &=& N_\pi \frac13 (C_1+C_2) \big[ 4\zeta^{B\pi} + 
  (4+\langle {\bar u}^{-1} \rangle_\pi)
\zeta_J^{B\pi}\big],\nn \\
T_c &=& N_\pi  \left[
\left(C_1+\frac{C_2}{3}+C_4+\frac{C_3}{3}\right) \zeta^{B\pi}\right. \\
&+& \left.
\left(C_1+C_4 + (1+ \langle {\bar u}^{-1} \rangle_\pi )\frac{C_2+C_3}{3}\right) 
\zeta_J^{B\pi} \right],\nn 
\end{eqnarray}
where $\langle {\bar u}^{-1} \rangle_\pi=\int_0^1\phi_\pi(u)/(1-u)$, and
$\zeta_J^{B\pi}=\int\! dz\, \zeta_J^{B\pi}(z)$.  The penguin amplitude also gets
a contribution from the complex $A^{\pi\pi}_{c\bar c}$ amplitude, so
\begin{eqnarray} \label{P}
P &\equiv& - \Big|\frac{\lambda_u^{(d)}}{\lambda_c^{(d)}}\Big|\,
  T_c\: r_c\, e^{i\delta_c} =
N_\pi\left[
\left(C_4+\frac{C_3}{3}\right) \zeta^{B\pi}\right. \nn \\
&&\hspace{-0.5cm} +
\left. 
\left(C_4 \!+\! (1\!+\! \langle {\bar u}^{-1} \rangle_\pi )\frac{C_3}{3}\right)
  \zeta_J^{B\pi} 
+\frac{1}{N_\pi} A_{c\bar c}^{\pi\pi}
\right]  \,.
\end{eqnarray}
The amplitude $T_n$ is given by the isospin relation Eq.~(\ref{isospin}) as $T_n
= T - T_c$. At tree level in SCET Wilson coefficients the $B\to \pi$ form factor
at $q^2=0$ is
\begin{eqnarray}\label{fplus}
  f_+(0)=\zeta^{B\pi}+\zeta_J^{B\pi}\,.
\end{eqnarray}
Neglecting the $O(\alpha_s(m_b))$ corrections introduces an error of about 10\%
for the $T$ amplitudes, which is smaller than the expected size of the power
corrections $\sim O(\Lambda/E)$.  

Eq.~(\ref{T}) implies that the tree amplitudes $T,T_c$ are calculable in terms
of the $\zeta, \zeta_J$ parameters, and their relative strong phase are small
$\theta, \theta_n \sim O(\alpha_s(m_b), \Lambda/E)$. On the other hand, the
penguin amplitude $P$ can have an O(1) strong phase due to the charming penguin
amplitude $A_{c\bar c}^{\pi\pi}$.
\begin{figure}[t!]
  \centerline{ \mbox{\epsfysize=6truecm \hbox{\epsfbox{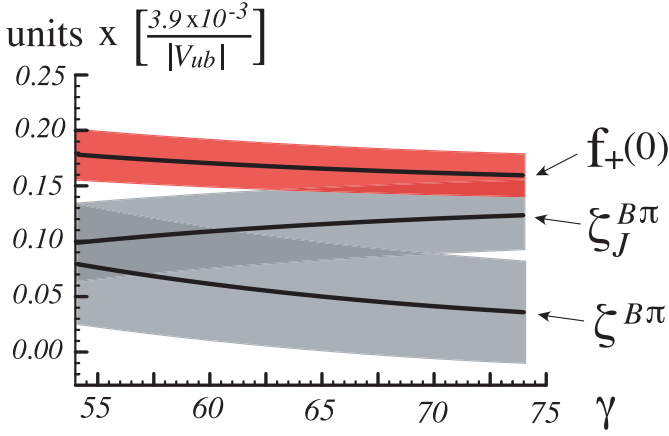}} } }
  \vskip-0.2cm {\caption[1]{Model independent results for $\zeta^{B\pi}$,
      $\zeta_J^{B\pi}$, and the $B\to \pi$ form factor $f_+(q^2=0)$ as a
      function of $\gamma$. The shaded bands show the $1$-$\sigma$ errors
      propagated from the $B\to\pi\pi$ data.}
\label{fig:ff} }
\vskip -0.4cm
\end{figure}
The pattern of results in Fig.~\ref{fig:pipidata} supports these predictions for
the tree amplitudes $T,T_c$ for the upper hand solution. In particular, within
the experimental uncertainty the phases $\theta$ and $\theta_n$ are still
consistent with being small and compatible with  order $O(\Lambda/E)$ effects.

Using the numbers in Eq.~(\ref{Texp}) for $|T|$ and $|t|$ and the SCET results
in Eqs.~(\ref{T}) we can extract the nonperturbative parameters $\zeta,\zeta_J$.
Taking LL order for the coefficients ($C_1=1.107, C_2=-0.248, C_3=0.011,
C_4=-0.025$ at $\mu=4.8 \mbox{GeV}$) and $\langle
\bar{u}^{-1}\rangle_\pi=3$~\cite{bakulev}, we find
\begin{eqnarray}\label{zetaexp}
 \zeta^{B\pi} \,\big|_{\gamma=64^\circ} & =& \big(0.05 \pm 0.05\big) 
      \Big( \frac{3.9\times 10^{-3}}{|V_{ub}|} \Big) 
      \,,\\
  \zeta_J^{B\pi} \, \big|_{\gamma=64^\circ} &=& \big( 0.11 \pm 0.03 \big) 
      \Big( \frac{3.9\times 10^{-3}}{|V_{ub}|} \Big)
     \,, \nn
\end{eqnarray}
where the quoted errors are propagated from the experimental errors from $|T|$
and $|t|$ in Eq.~(\ref{Texp}).  Using the results for $r_c$ and $\delta_c$ in
Eq.~(\ref{rcdelc}) and $|V_{cb}|=0.041$ the penguin amplitude is
\begin{eqnarray} \label{Pnum}
 \frac{P}{N_\pi}  \Big|_{\gamma=64^\circ}
  = (0.043\pm 0.013) \,
  e^{i( 136^\circ \pm 12^\circ) } 
 \,.
\end{eqnarray} 
%\begin{eqnarray}
% \frac{1}{N_\pi} A_{c \bar c}^{\pi\pi} \big|_{\gamma=64^\circ}
%  = (0.042\pm 0.013) \,
%  e^{i( 134^\circ \pm 12^\circ) } 
% \,.
%\end{eqnarray} 
The $\zeta^{\pi\pi}$ and $\zeta_J^{\pi\pi}$ terms in Eq.~(\ref{P}) contribute
$0.002$ to $P/N_\pi$, which is only a small part of the experimental result.
The perturbative corrections from the $\Delta c_i^{(f)}$'s or particularly the
$\Delta b_i^{(f)}$'s can add terms whose rough size is estimated to be $\sim
\zeta_J^{B\pi} C_1\,\alpha_s(m_b)/\pi\simeq 0.007$. After removing these
contributions, the sizeable remainder would be attributed to $A_{c\bar
  c}^{\pi\pi}$. Since $A_{c\bar c}^{\pi\pi}$ can have a large non-perturbative
strong phase, the large phase in Eq.~(\ref{Pnum}) supports the conclusion that
this term contributes a substantial amount to $P/N_\pi$.

The extraction of the above parameters allows us to make two model independent
predictions with only $\gamma$ and $|V_{ub}|$ as input. First a prediction for
the semileptonic $B\to \pi$ form factor $f_+(0)$ is possible.  Combining
Eq.~(\ref{zetaexp}) with Eq.~(\ref{fplus}) we find
\begin{eqnarray} \label{f+0}
 f_+(0) \, \big|_{\gamma=64^\circ}
  = \big(0.17\pm 0.02\big)  \Big( \frac{3.9\times 10^{-3}}{|V_{ub}|} 
  \Big) \,.
\end{eqnarray}
In Fig.~\ref{fig:ff} we show results for $\zeta^{B\pi}$, $\zeta_J^{B\pi}$, and
$f_+(0)$ for other values of $\gamma$, thus generalizing the results in
Eqs.~(\ref{zetaexp}) and (\ref{f+0}). Note that including the correlation in the
errors for $\zeta^{B\pi}$ and $\zeta_J^{B\pi}$ has led to a smaller uncertainty
for $f_+(0)$.  Theory uncertainty is not shown in Eq.~(\ref{f+0}) or
Fig.~\ref{fig:ff}, and the most important source are power corrections which we
estimate to be $\pm 0.03$ on $f_+(0)$. One loop $\alpha_s(m_b)$ corrections
are also not yet included. Varying $\mu=2.4$--$9.6\,{\rm GeV}$ in the LL
coefficients changes $f_+(0)$ by only a small amount $\mp 0.01$.

It is interesting to note that the central values from our fit to the data give
$\zeta_J^{B\pi}\gtrsim \zeta^{B\pi}$ which differs from the hierarchy used in
QCDF.  Furthermore our central value for $f_+(0)$ is substantially smaller than
the central values obtained from both QCD sum rules~\cite{Ball} ($f_+(0)=0.26$),
from form factor model based fits to the semileptonic data~\cite{Luo}
($f_+(0)=0.21$), or those used in the QCDF analysis~\cite{BBNS2} ($f_+(0)=0.28$
or $0.25$).

Our analysis can also be used to make a prediction for $Br(B^0\to \pi^0\pi^0)$.
At tree level in SCET  $|t_n|=|t|-1$ which gives
\begin{eqnarray} \label{B00th}
   \frac{\bar \Gamma(B^0\to \pi^0\pi^0)}{\bar\Gamma(B^-\to \pi^0\pi^-)} 
  \!\!&&\!\!  \\
  && \hspace{-2.2cm} 
  =\Big(\frac{|t|\!-\!1}{|t|}\Big)^2 + \frac{r_c^2}{|t|^2}
  - \frac{2 r_c}{|t|} 
  \Big(1\!-\!\frac{1}{|t|}\Big)\cos(\delta_c) \cos(\gamma)\,. \nn
\end{eqnarray}
Thus we predict
\begin{eqnarray} \label{B00}
  Br(B^0\!\to\! \pi^0\pi^0) =\! \left\{ \!
  \begin{array}{cc}
    (1.0 \pm 0.7)\! \times\!  10^{-6}, & \gamma=54^\circ \\
    (1.3 \pm 0.6)\! \times\!  10^{-6}, & \gamma=64^\circ \\
    (1.8 \pm 0.7)\!  \times\! 10^{-6}, &\gamma=74^\circ
   \end{array} \right. \!\! .
\end{eqnarray}
These results are all in reasonable agreement with the current world average.
The uncertainty quoted in Eq.~(\ref{B00}) is only from the inputs in
Eq.~(\ref{B00th}), and will be directly reduced when the first four measurements
in Eq.(\ref{pipidata}) improve.  Since the $\zeta_J^{B\pi}$ term in
Eq.~(\ref{zetaexp}) is $\gtrsim \zeta^{B\pi}$ our results for $Br(B^0\to
\pi^0\pi^0)$ are not contaminated and we expect that theoretical uncertainty
from power corrections plus $\alpha_s(m_b)$ corrections will add a $\sim
20$-$30\%$ uncertainty to the results in Eq.~(\ref{B00}).  Note that one can
turn the analysis in Eq.~(\ref{B00}) around and use the data on $B \to \pi \pi$
in Eq.~(\ref{pipidata}) to give a new method for determining the value of
$\gamma$, where the theoretical input from factorization is that the tree
triangle is flat.

Our values in Eq.~(\ref{B00}) are somewhat larger than the central values
predicted in QCDF ($\sim 0.3\times 10^{-6}$~\cite{BBNS2}) or pQCD ($\sim
0.2\times 10^{-6}$~\cite{Lu}).  For $\gamma=54^\circ$ the first term in
Eq.~(\ref{B00th}) dominates our result, while the $r_c^2$ penguin term has a
large cancellation with the interference term $\propto \cos(\gamma)$. For larger
$\gamma$'s this cancellation becomes less effective and $Br(B^0\to\pi^0\pi^0)$
increases.  In QCDF $\zeta^{B\pi}$ dominates over a small $\zeta_J^{B\pi}$, but
has a small coefficient $\propto C_2+C_1/3$, so the first term in
Eq.~(\ref{B00th}) is small.  In pQCD the $M_{a,e}$ terms which are multiplied by
$C_1$ are also small for $B\to \pi^0\pi^0$.

In this paper we have used SCET to derive a factorization theorem for $B\to M_1
M_2$ decays and explored the theoretical and phenomenological implications.
Several issues for $B\to M_1 M_2$ still remain to be resolved.  A factorization
formula for the charming penguin contribution should be worked out, and
polarization effects should investigated beyond leading order.  It needs to be
shown that the $n$--$\bar n$ factorization is not spoiled by Glauber degrees of
freedom.  The one loop $\Delta b_i$'s need to be computed, as well as a
resummation of Sudakov logarithms which are given by the evolution equations for
the SCET operators. Charming penguin effects need to be better understood in an
effective theory approach, and a full factorization theorem for the $A_{c\bar
  c}$ amplitude should be worked out.  Finally, power corrections (including so
called chirally enhanced terms, annihilation contributions, and $C_1\Lambda/E$
terms) should be studied using SCET.

This work was supported in part by the DOE under DE-FG03-92ER40701,
DOE-ER-40682-143, DEAC02-6CH03000, and the cooperative research agreement
DF-FC02-94ER40818. I.S. was also supported by a DOE OJI award.

\vspace{-.4cm}
%\newpage

\newpage

\end{document}